# Metaverse: A Vision, Architectural Elements, and Future Directions for Scalable and Realtime Virtual Worlds

Leila Ismail[*,1,2] and Rajkumar Buyya[1]

[1] Cloud Computing and Distributed Systems (CLOUDS) Lab
School of Computing and Information Systems
The University of Melbourne, Australia

[2] Intelligent Distributed Computing and Systems (INDUCE) Lab
Department of Computer Science and Software Engineering
National Water and Energy Center
United Arab Emirates University, United Arab Emirates

**Abstract**
With the emergence of Cloud computing, Internet of Things-enabled Human-Computer Interfaces, Generative Artificial Intelligence, and high-accurate Machine and Deep-learning recognition and predictive models, along with the Post Covid-19 proliferation of social networking, and remote communications, the Metaverse gained a lot of popularity. Metaverse has the prospective to extend the physical world using virtual and augmented reality so the users can interact seamlessly with the real and virtual worlds using avatars and holograms. It has the potential to impact people in the way they interact on social media, collaborate in their work, perform marketing and business, teach, learn, and even access personalized healthcare. Several works in the literature examine Metaverse in terms of hardware wearable devices, and virtual reality gaming applications. However, the requirements of realizing the Metaverse in realtime and at a large-scale need yet to be examined for the technology to be usable. To address this limitation, this paper presents the temporal evolution of Metaverse definitions and captures its evolving requirements. Consequently, we provide insights into Metaverse requirements. In addition to enabling technologies, we lay out architectural elements for scalable, reliable, and efficient Metaverse systems, and a classification of existing Metaverse applications along with proposing required future research directions.

## 1. Introduction

Advances in information and computing technologies, such as the continuous availability of high-speed 5G and 6G networks to Internet users, Internet of Things enabled Human-Computer Interfaces, high precision of data-driven Machine and Deep Learning models, and the emergence of Generative Artificial Intelligence with ChatGPT, have enabled the rise of Metaverse systems and applications.

The term Metaverse was first coined by Neal Stephenson in his science fiction novel "Snow Crash" in 1992 to describe a future virtual reality [1], has recently emerged as the *next Internet revolution*, providing information and communication facilities to a network of a virtual world, with the possibility to create and interact with a virtual space whereas disparate Internet of Things (IoT) and Internet of Humans communicate in a spatial environment. Metaverse is different from VR and AR Virtual Reality (VR) [2] and Augmented Reality (AR) [3] in the sense that Metaverse offers services that enable people to interact with each other and with the virtual environment via their avatars, who can shop, socialize, play interactive games, work, learn, do tourism, or consult a medical doctor virtually but has the sensation of being physically present. While VR and AR aim to create a continuous flow of 3D sensory images that represent a physical world, Metaverse offers services that have sustainable content. Despite that the definition of "Metaverse" has changed as the integrating technology evolved, the main goal is to have a digital twin of the physical world, while providing a realtime fully immersive virtual 3D space where the physical and the virtual world can interact, providing efficient, safe, and pleasant experiences in all domains, such as smart education, smart healthcare, smart transportation, and geospatial localizations. A radical evolution of the computing era where the Internet of Everything is transformed into a User-Centric Intelligent Internet of Everything. Metaverse has stringent requirements in terms of high data rate, high reliability, low latency, and connected intelligence with Machine Learning (ML) and Deep Learning (DL). The upcoming Sixth Generation (6G) of wireless network technology is designed to support efficient, dependable, and secure Metaverse applications in the smart city while considering the critical requirements of privacy, energy efficiency, high data rates, and ultra-low latencies of those applications [4].

---
[*] Corresponding author (Ismail); email: leila@uaeu.ac.ae

Metaverse presents a massive research opportunity. It is expected that the global market for Metaverse technology will reach almost $679 billion in 2030, from over $47 billion in 2022. This is an unprecedented annual growth of more than 39% [5]. Gartner considers Metaverse as one of the top 10 strategic technology trends for 2023 [6] and predicted that by 2026, 25% of people will spend at least one hour a day on work, shopping, education, social, and/or entertainment [7]. Consequently, Facebook has rebranded into Meta Platform [8] and has been developing Metaverse applications, such as Horizon Workrooms [9] and Horizon World [10], and Microsoft showed increased interest in pioneering Metaverse by investing in Artificial Intelligence (AI)-enabled telehealth care, Interactive Voice Response (IVR) and virtual assistants [11].

This paper presents the current trends in Metaverse research driven by applications and the need for convergence in several interdisciplinary technologies. The rest of the paper is organized as follows. Section 2 compares related work. In Section 3, we present Metaverse trends and a retrospective analysis of the Metaverse temporal evolution through its definitions in terms of requirements. Section 4 discusses the Metaverse's overall vision and the enabling technologies. We present a layered Metaverse architecture and explain its elements in Section 5. We compare the existing Metaverse development platforms and discuss their advantages and limitations for the realization of Metaverse applications in Section 6, followed by a taxonomy of Metaverse applications in Section 7. We discuss open challenges and propose future directions in Section 8 and conclude in Section 9.

## 2. Related Works

There have been few surveys on Metaverse [12]–[15]. We categorize these surveys based on their focus into 3 categories, 1) definitions of Metaverse [12], 2) requirements [13], and 3) Metaverse enabling technologies, applications, and challenges [14], [15]. [12] performed a systematic review of the literature to synthesize the definition of the Metaverse. The author concluded that Metaverse as an immersive, synchronous, and persistent virtual world (overlapped with the physical world) that allows users, represented by avatars, to interact with each other and the environment. [13] described the current status of Metaverse characteristics which are realism, ubiquity, interoperability, and scalability. [14] discussed the existing security and privacy threats in the Metaverse. The authors classified the threats into authentication and access control, data management, privacy-related, network-related, economy-related, physical/social effects, and governance-related categories. Similarly, [15] analyzed different security and privacy issues in the Metaverse and presented some possible countermeasures. Furthermore, they described Metaverse enabling technologies and application areas. Different from the above related surveys [12]–[15], we focus on a vision of achieving realtime and scalable Metaverse, its enabling technologies, requirements, an architecture perspective, challenges, and future directions.

In this paper, we present a comprehensive survey of the Metaverse key requirements, enabling technologies, as well as architecture and development platforms to build realtime and scalable Metaverse. By discussing existing challenges and potential solutions, this survey provides critical insights on how to build realtime and scalable software solutions for developing green and dependable Metaverse. The contributions of this paper are six-fold.

- We investigate the temporal evolution of Metaverse definitions to extract the key requirements (i.e., immersive and multisensory interaction, spatiotemporality, interoperability, scalability, heterogeneity, QoS (Quality of Service), and QoE (Quality of Experience)), which are fundamental to developing a realtime and scalable Metaverse by analyzing the temporal evolution of Metaverse definitions.
- We discuss the components of Metaverse (i.e., environment, interface, interaction, and data security and privacy), and their enabling technologies (i.e., generative AI, deep learning and machine learning, IoT, blockchain, edge and cloud computing, digital twins, VR, AR, spatial computing, computer vision, web3, and network), to enable users to have an immersive, multisensory, interactive, and secure experience.
- We propose a layered architecture, underpinned by the enabling technologies, toward building green and dependable Metaverse distributed computing applications. We divide the architecture into four decoupled layers (infrastructure, distributed computing, platform, and application), where each layer can evolve independently, to ensure the Metaverse applications requirements.
- We survey the development platforms for Metaverse applications and compare them to provide insights and useful guidelines for developers into building domain specific Metaverse applications.
- We provide a taxonomy of Metaverse applications based on their domains of development (i.e., education, smart healthcare, smart mobility, gaming and entertainment, business, social media, and manufacturing), and categorize the existing Metaverse application accordingly.
- We discuss the critical challenges for Metaverse realization (i.e., realtime, scalability, high energy consumption, resource provisioning to optimize QoS and energy consumption, cost and complexity, security and data privacy, the need for governance against abuse, standardization and interoperability, as well as health-related risks), and outline possible solutions for future research directions.

Table 1 summarizes the contribution of our work in comparison to the previous related surveys.

**Table 1:** Summary of related surveys.

| Survey | Definitions | Requirements | Components | Enabling technologies | platforms | Taxonomy of applications | Architecture | Open issues and possible solutions |
|---|---|---|---|---|---|---|---|---|
| [12] | ✓ | ✗ | ✗ | ✗ | ✗ | ✗ | ✗ | ✗ |
| [13] | ✗ | ✓ | ✗ | ✗ | ✗ | ✗ | ✗ | ✗ |
| [14] | ✗ | ✓ | ✗ | ✓ | ✗ | ✓ | ✗ | ✓ |
| [15] | ✗ | ✗ | ✗ | ✓ | ✗ | ✓ | ✗ | ✓ |
| This survey | ✓ | ✓ | ✓ | ✓ | ✓ | ✓ | ✓ | ✓ |

## 3. Metaverse Trends, Definitions, and Requirements

### 3.1 Trends

According to a Gartner report [6], by 2027, over 40% of large organizations worldwide will be using a combination of Web3, spatial computing, and digital twins in Metaverse-based projects aimed at increasing revenue. These technologies are enabled by IoT, edge, and cloud computing integrated systems to satisfy the stringent requirements of the Metaverse realtime applications in terms of QoS and respecting a Service-Level Agreement (SLA). We envision that Metaverse will be rapidly and widely adopted thanks to the projection of being a vendor-independent/portable computing platform. It will have a great impact on the economy, as Metaverse will increase the potential customers' QoE, and enable rapid, and realtime interactions with users. In addition, it will have its digital economy enabled by blockchain [16], where users can sell and buy their digital assets using digital currency.

Metaverse is pointed out as one of the 2023 emerging technologies trends, as shown in Gartner's radar (Figure 1). The radar [6] shows the stages over time of each emerging technology from early adoption to majority adoption by applications. The range in the radar measures the number of years it will take the technology to cross over from emergence to maturity. The mass represents how substantial the impact of the technologies will be on existing products and markets. It is projected that the Metaverse will take 6 to 8 years for market adoption. As shown in Figure 2, during the last 5 years the search popularity of Metaverse has been increasing, with a very sharp increase in January 2021 due to the announcement of Facebook to change its name to Meta to realize the vision of Metaverse, but since then, VR has been falling behind [17]. This is thanks to the rapid technological advancement of VR and its wearable devices [18] (Figure 3), which are integrating parts of the Metaverse, so the latter took over VR. Wearable devices include VR Metaverse devices (such as Meta Quest 2 by Reality Labs in Meta, Valve Index, Sony PlayStation VR, HP Reverb G2 by Hewlett Packard, and Haptx Gloves) and AR Metaverse devices (such as Microsoft Hololens, Magic Leap, Mojo Vision, Epson Moverio BT-350, and Google Glass Edition 2) [19], [20].

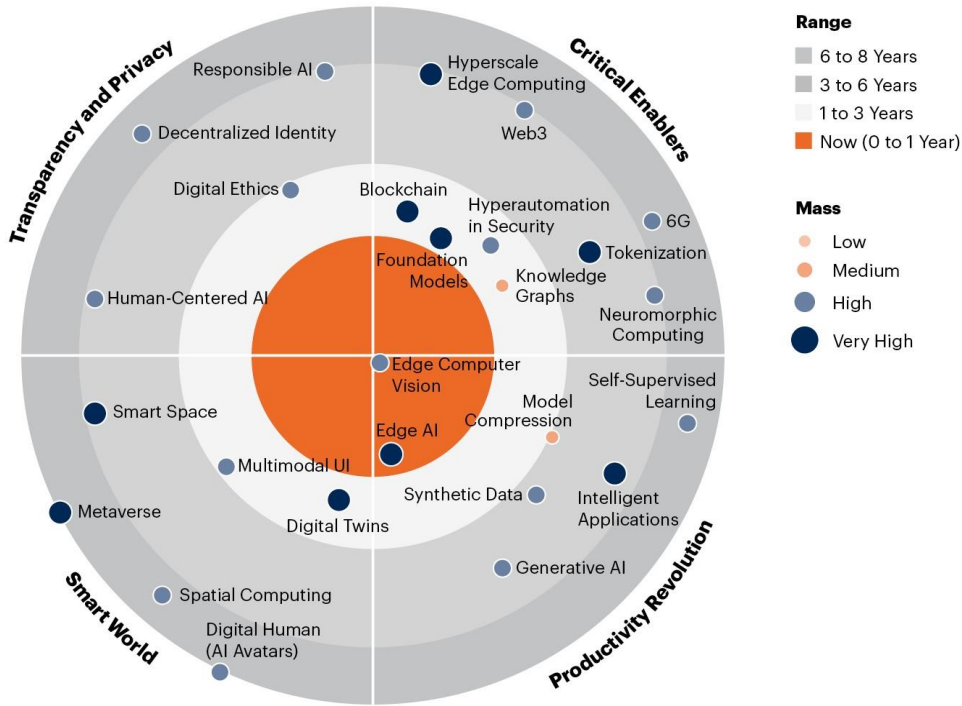

**Figure 1:** Emerging technologies trends (Source: Gartner 2023 [6]).

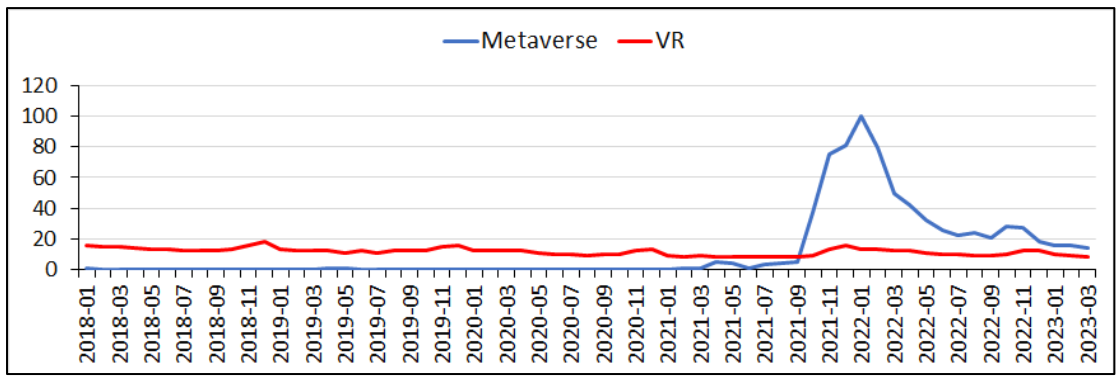

**Figure 2:** Web search trends worldwide since 2018 for the terms "Metaverse" and "Virtual Reality (VR)".

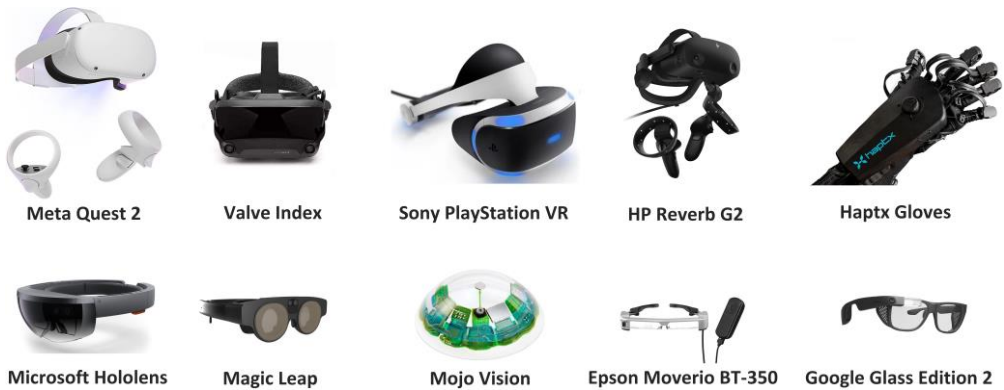

**Figure 3:** Metaverse wearable devices.

## 3.2 Definition, Evolution, and Requirements of Metaverse

Over the last two decades, Metaverse has undergone quite a few reformations/transformations, alongside technological advances, to address the need for a more reliable, seamless, and immersive experience and accordingly many definitions have evolved. To extract these definitions, we reviewed published articles and reports on Metaverse. This is by searching relevant literature in the Association for Computing Machinery (ACM), Elsevier, Institute of Electrical and Electronics Engineers (IEEE), MEDLINE, PubMed, Scopus, and Web of Science databases. Table 2 presents the evolving definitions of the Metaverse.

**Table 2:** The definitions of Metaverse.

| Number | Year | Source | Definition |
|---|---|---|---|
| 1 | 1996 | [21] | "a future version of the Internet which appears to its participants as a *quasi − physical world*. Participants are represented by *fully articulate human figures, or avatars*. Body movements of avatars are computed automatically by the system" |
| 2 | 1998 | [22] | "a *virtual reality world* envisioned as a large cyber-planet. It contains *homes, corporate headquarters, nightclubs, and virtually every other type of building* found in reality and some that are not. Individuals from around the world materialize on this cyber-planet, and are represented there by *avatars*" |
| 3 | 2007 | [23] | "a virtual world which is a genre of online community that often takes the form of a computer-based *simulated environment*, through which users can *interact* with one another and use and create objects" |
| 4 | 2008 | [24] | "an extensive 3D networked virtual world capable of supporting a *large number of people simultaneously* for social interaction […] implies the interaction of real people with the virtual environments and agents including avatars with increasing levels of *immersion and presence* […] the word metaverse (Meta -Universe) suggests the emergence of a new class of augmented *social interaction* which we term 'augmented duality'" |
| 5 | 2008 | [25] | "a system of *numerous*, interconnected *virtual* and typically *user-generated worlds* (or Metaworlds) all *accessible through a single-user interface*" |
| 6 | 2010 | [26] | "a user can walk around *realistically* through an *avatar* […] The most realistic interface is a 3D virtual world, also known as a 'metaverse' " |
| 7 | 2013 | [13] | "refers to a fully *immersive three dimensional digital environment* in contrast to the more inclusive concept of cyber space that reflects the totality of shared online space across all dimensions of representation" |
| 8 | 2013 | [13] | "an integrated *network of 3D virtual worlds*" |
| 9 | 2021 | [27] | "a ***3D-based virtual reality*** in which daily activities and economic life are conducted through ***avatars*** representing the real themselves" |
| 10 | 2021 | [27] | "metaverse means a world in which virtual and reality ***interact*** and co-***evolve***, and social, economic, and cultural activities are carried out in it to ***create value***." |
| 11 | 2021 | [28] | "an evolving virtual world with unlimited ***scalability*** and ***interoperability***. The operators need to construct the basic elements, while innovative user-generated content (UGC) fulfill the universe through users. Therefore, ***high efficiency*** content creation is another significant component for interactions between users and the metaverse." |
| 12 | 2021 | [29] | "the next evolution in ***social connection*** and the successor to the ***mobile internet***" |
| 13 | 2022 | [30] | "refers to a ***created world***, in which ***people can live*** under the rules defined by the creator" |
| 14 | 2022 | [31] | "is a ***massively scaled*** and ***interoperable network*** of ***real-time rendered 3D virtual worlds*** which can be experienced synchronously and persistently by an ***effectively unlimited number of users*** with an individual ***sense of presence***, and with continuity of data, such as ***identity, history, entitlements, objects, communications, and payments***" |
| 15 | 2022 | [32] | "is the post-reality universe, a perpetual and persistent ***multiuser environment merging physical reality with digital virtuality*** […] enable ***multisensory interactions*** with virtual environments, digital objects and people such as virtual reality (VR) and augmented reality (AR) […] is an interconnected web of social, ***networked immersive environments in persistent multiuser platforms***. It enables s***eamless embodied user communication*** in ***real-time*** and ***dynamic interactions*** with digital artifacts" |
| 16 | 2023 | [33] | "it can be something that transcends physical reality described in terms of ***time and space*** […] It can denote a universe distinct from the physical universe but referring to it by summarizing, condensing, or depicting its various aspects […] It can refer to ***one or more potential possible alternatives to the existing universe***" |

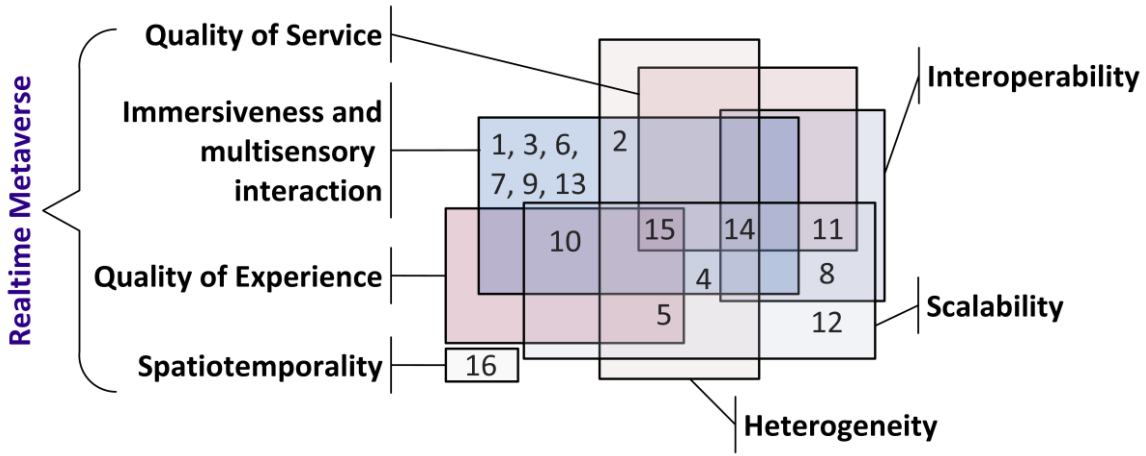

**Figure 4:** Requirements of Metaverse (*Note: The numbers represent the definition number from Table 2*).

Based on these definitions, we provide a retrospective analysis of the Metaverse requirements which evolved over time. These requirements are as follows (Figure 4):

- *Immersive and multisensory interaction:* Metaverse should allow users to feel and experience emotional as well as psychosocial involvement by rendering a realistic virtual environment [26]. This is achieved through sensory perceptions (e.g., temperature, sound, touch, and sight) and expressions (for instance, gestures). Sensory images need to be produced fast enough for users to perceive them as a continuous flow rather than discrete events. Realtime rendering of Metaverse is crucial to the success of Metaverse, aiming at increasing the Metaverse QoS while respecting SLAs requirements and consequently improving users' QoE.
- *Spatiotemporality:* Metaverse should allow users to freely navigate across different digital worlds with disparate spatiotemporal dimensions which is not possible in the real physical world due to the finiteness of space and irreversibility of time [14]. It should break these boundaries of space and time.
- *Interoperability:* Metaverse should enable interoperability between digital assets and data, collected using different Metaverse wearable devices and IoT devices, disparate virtual spaces implemented using different platforms, for the rendering of a virtual world [13]. Consequently, users can seamlessly shuttle across different virtual worlds without interruption.
- *Scalability:* Metaverse should perform efficiently with the increasing number of simultaneous users or avatars, scene complexity, and interactions between users from different virtual worlds [13]. It should scale up automatically, supporting the increasing number of connections, processing, and Input/Output operations, to support realtime interactions and rendering the virtual worlds, without affecting the QoS and the users' QoE.
- *Heterogeneity:* Metaverse should support the use of heterogeneous wearable devices (e.g., hand-based, non-hand-based, and head-mounted), heterogenous data types (e.g., structured or non-structured, text or image), different communication modes and network protocols (e.g., cellular, Wi-Fi, 5G, and 6G), and diverse human perception, behavior, and psychology [34].
- *Quality of Service:* Metaverse should satisfy QoS requirements in terms of ultra-low latency, high bandwidth, and high data rate for realtime rendering of complex scenarios for multiple users. The 5G and its next-generation 6G networks consist of promising ultra-rapid communication protocols in a highly heterogeneous and dynamic setup such as the Metaverse [4].
- *Quality of Experience:* Metaverse should provide a seamless and uninterrupted experience to the users by ensuring there is no lag/delay in the interaction between the physical and virtual world, high visual quality, realtime tactile, and control experience [35]. The users should have an effective and immersive virtual encounter using an easy-to-use interface.

## 4. Enabling Technologies for Metaverse Computing

The Metaverse is a new computing era where the Internet becomes a shared, persistent, and immersive 3D virtual dynamic, open, and interoperable space, where people, avatars, Robotics, and the Internet of Things can interact as if they are in the physical world. Figure 5 depicts Metaverse components, which comprise four basic components: 1) environment, 2) interface, 3) interaction, and 4) data security and privacy [36]. The environment component is necessary to identify the sights and sounds from the physical world to design and render a digital world. The visual environment can be composed by recognizing and rendering objects and scenes, whereas the sound environment can be synthesized using sound and speech recognition. Furthermore, the movement of avatars in a virtual world is facilitated by motion rendering. The interface component enables users to have an immersive and multisensory experience.

The interaction component aids in bi-directional interactivity between the physical and virtual worlds via multimodal interaction. In addition, multi-request interaction allows avatars to perform multiple tasks simultaneously such as motion, conversation, and sensing. This component enables 3D interaction and human behaviour modelling to enrich the overall experience. The security and privacy component ensures the integrity and privacy of users' data confidential data such as biometrics information and personal identities. Furthermore, it provides security for the communication network, wearable devices, and underlying software.

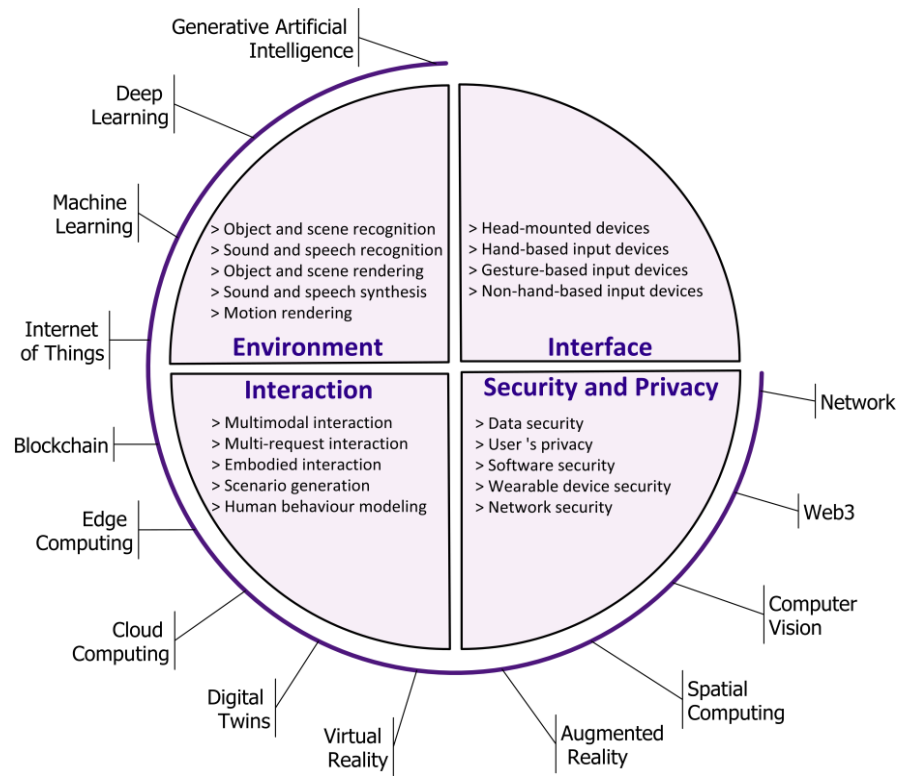

**Figure 5:** Metaverse: Enabling Technologies

In the following, we explain different enabling technologies for the Metaverse (Figure 5).
- *Generative Artificial Intelligence:* It uses learning models for examining patterns in existing data to generate new content. Generative AI will enable users to customize their virtual environments for a more personalized and engaging experience. It will make the Metaverse more accessible by providing tools for bi-directional text and speech translations, accommodating different languages and cultural backgrounds, as well as assisting differently-abled users while promoting equity, diversity, and inclusion.
- *Deep Learning and Machine Learning:* For the Metaverse to provide an immersive and personalized experience to avatars and users, it should support context awareness, which relies on smart observations of information related to avatars such as their locations. For instance, to provide recommendations on nearby virtual places to visit or create a suitable virtual environment for a patient knowing health status and conditions (such as, such as a particular disability) to provide comfortable user experiences. This can be realized using the power of the machine and deep learning by developing context-aware predictive models.
- *Internet of Things:* IoT devices and sensors are used to generate data in the physical world for the creation of a virtual world. Context-aware data generated from a network of IoT sensors and devices would lead to different decision-making approaches. For instance, conditions under which the data has been generated, e.g., heart acceleration for a running user versus sitting.
- *Blockchain:* It is a privacy-preserving enabler for data and associated transactions generated in the virtual world [16] [37]  Furthermore, for Metaverse applications that involve financial transactions, a Blockchain-based marketplace can be used for buying and selling virtual good known as Non-Fungible Tokens (NFTs) (art, music, videos, photography, trading cards, etc.) [38]. Transactions in the virtual worlds can be tracked via a persistent and immutable distributed ledger.
- *Edge and Cloud Computing:* It will be a crucial part of the Metaverse providing an integrated computing system utility for the virtual platform. Cloud is highly heterogeneous and provides highly scalable storage and computing capabilities [39] for the Metaverse applications, with various tools to develop models for reactive and/or predictive data analytics to support a fully immersive and interactive virtual world, based on data generated from the physical world. Edge data centers, close to users, enable low-latency and realtime computing and inferences to Metaverse applications and avatars.
- *Digital Twins:* It is a digital replica of a physical entity or object that is continuously updated with the performance and maintenance data of the physical system [40], [41]. It will aid in determining and predicting the behavior of any physical entity in the Metaverse environment.
- *Virtual Reality:*  It is an immersive experience where the physical space is replaced by a computer-simulated digital environment with realtime interaction [2]. A user interacts with the digital world using input devices such as hand-held controllers and haptic gloves. VR includes the digital world and excludes physical space.
- *Augmented Reality:* It supplements the physical world, instead of replacing it as in VR, by digitally superimposing digital information and objects on physical objects for simultaneous interactions with both virtual and physical spaces [3]. AR does not allow interactions with digitally imposed data. In Enhanced AR, known as Mixed Reality, the physical world interacts with the superimposed digital data in realtime [42]. MR mitigates the limitations of both VR and AR by including the physical world and the ability to interact with the digital world.

- *Spatial Computing:* It is a form of human-computer interaction that retains and controls objects in the physical world. Spatial computing in the Metaverse will enable an immersive and interactive experience for users that will diminish the line of difference between the physical and virtual worlds [32].
- *Computer Vision:* It refers to the class of AI that develops learning models for visual data such as images and videos. In the Metaverse, computer vision will be used to track users in the physical space and represent them as avatars in the virtual world [43].
- *Web3:* It refers to the 3$^{rd}$ generation of the web to realize the vision of a decentralized web, web1 is static content, web2 is dynamic content and Social Media features, and web3 advocates more decentralization so that users are in control of their data, originally called the Semantic Web by inventor the Semantic Web Tim Berners-Lee [44], allow to process information more intelligently through Big Data, Machine Learning, and Decentralized Blockchain ledger technology [45].
- *Network:* Metaverse applications require ultra-reliability and a high data rate for the wireless system to ensure Quality of Experience. The previous generation 4G network is not capable of providing low latency and high reliability for the Metaverse applications. The Enhanced Mobile Broadband (eMBB), Massive Machine-type Communications (mMTC), and Ultra-Reliable Low-Latency Communications (URLLCs) services of 5G networks can support VR and AR applications by provisioning large bandwidth and low latency with a certain level of reliability. However, 5G might degrade QoE for dynamic, interactive, and immersive applications. The Computation Oriented Communications (COC), Contextually Agile eMBB Communications (CAeC), and Event-Defined Ultra-Reliable Low-Latency Communications (EDuRLLC) application services of the 6G network [4] will ensure the stringent visual and physical requirements, of high bandwidth and low latency communication, for an immersive Metaverse experience.

## 5. Metaverse: An Architecture Perspective

Figure 6 represents an overview of Metaverse layered architecture that enables an immersive experience for users in different application domains such as smart healthcare, smart mobility, smart education, business, manufacturing, gaming and entertainment, and social media. We divide the architecture into four layers: infrastructure, distributed computing, platform, and application.

The infrastructure layer consists of IoT sensors and Metaverse wearable devices, edge and cloud computing systems, storage, and networking components. IoT sensors aid in collecting data from the physical world to construct a virtual world, whereas wearable devices provide an immersive experience to the user through a simulated avatar that replicates a user's natural reactions and emotions in a digital world. For instance, when playing an online game, the users' body language, heart rate, and breathing pattern will be replicated on the corresponding avatar creating an immersive sensation and increasing QoE. IoT enables the creation of an immersive virtual ecosystem. For instance, it enables the implementation of a virtual smart city Metaverse where avatars can participate in testing the development of the smart city ecosystem, such as testing vehicles in specific conditions, or the implementation of logistics deployments, such as fleets and traffic flow for urban planning and constructions. Head-mounted wearable devices

(such as Meta Quest 2 and Valve Index) show the virtual world (3D images and videos) on the device display, and a Beacon connects to mobile devices and provides content-local-based services to the user [20]. Global Market Insights predicted that the beacon technology market would surpass US$25 billion by 2024 [46]. Combined with HMD, beacons can provide content services to HMD so that the user's virtual world is provided with information of interest to users based on their locations. Connected to HMD via Bluetooth in a museum virtual tour, beacons exhibit contents to allow visitors better understand the artifacts on display, thus providing the grounds for building a Metaverse museum tour [47]. Metaverse wearable devices make use of the IoT, digital twins, VR, and AR technologies to satisfy the immersiveness and multisensory interaction requirement of the Metaverse.

The gateway devices, such as Raspberry Pi, mobile phones, and computers, enable data authentication, aggregation, and preprocessing. The edge servers are placed within routers and base stations in proximity to IoT devices, whereas the cloud servers are remote and geo-distributed. The virtual world is rendered in the cloud data centers as they have higher computing and storage capabilities than edge servers. However, cloud servers result in higher latency than edge servers, as they are situated far from the end users. The edge and cloud storage components store the ledger that contains data for different events such as the creation of Metaverse, simulation, optimization, prediction, monitoring, and controlling. The storage capabilities of cloud servers are higher than that of edge servers. The cloud and edge computing paradigms ensure the scalability requirement of the Metaverse. The network component provides communication of user behaviors and rendered virtual scenes among different components and end-users.

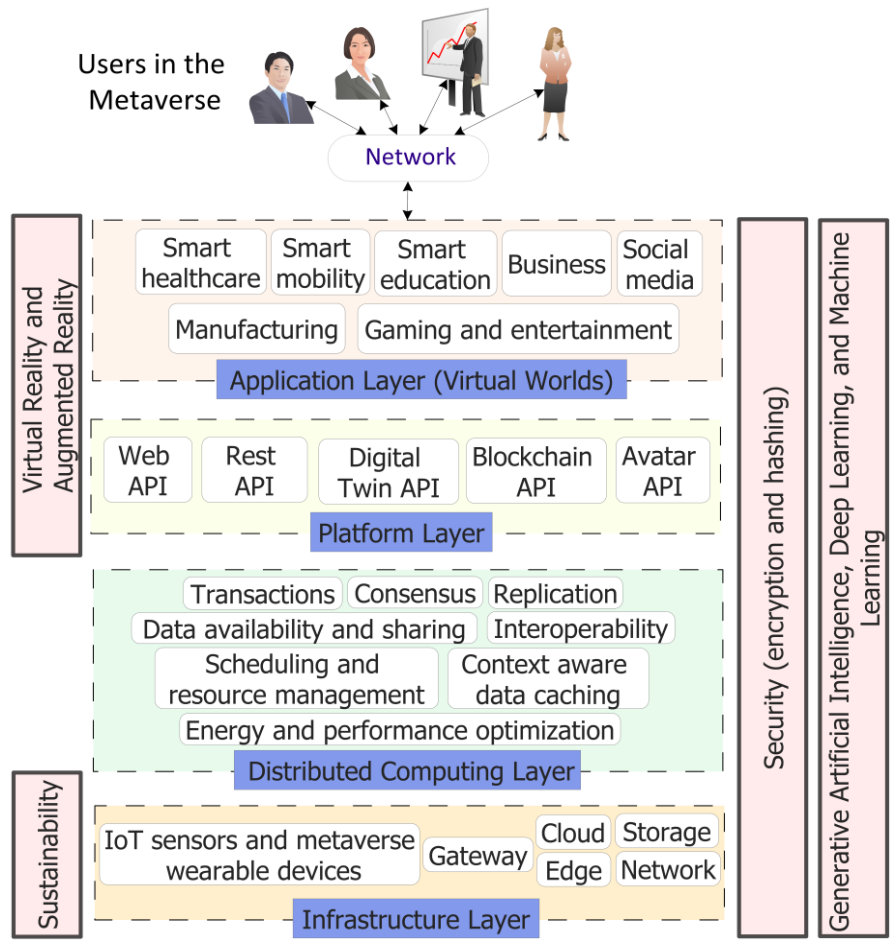

**Figure 6:** Overview of Metaverse layered architecture.

The distributed computing layer provides local access to data, replication for fault tolerance and availability, a consensus mechanism, and interoperability. Consensus protocol is used by the blockchain network to reach an agreement regarding data access, virtual world creation, and ledger update in a peer-to-peer manner. In addition, this layer uses AI approaches for data sharing, context-aware data caching, and energy-performance-aware resource management, scheduling, and communication. The platform layer enables Remote Procedure Calls (RPC) [16], web Application Programming Interface (API) [16], REpresentational State Transfer (REST) APIs [16], Digital Twin APIs, blockchain APIs, and avatar APIs for the creation of virtual worlds and communication between the network participants. The application developed for different domains such as smart healthcare, smart mobility, smart education, business, manufacturing, gaming, and social media can be accessed by users over a 6G network. The 6G network with high data rates and ultra-reliable low latency communication enables an immersive experience for the users in realtime.

The energy consumption of edge and cloud computing resources is increasing at an alarming rate, leading to high carbon emissions [48]. Consequently, the infrastructure layer uses AI approaches for energy-aware resource management, application communication, and execution. This ensures the sustainability of the underlying architecture and reduces global warming. VR and AR technologies are used across the platform and application layers to provide immersive and interactive experiences to users in virtual worlds. Furthermore, to provide security, blockchain technology is underpinned across all four layers for user authentication by encryption technique [49] and data integrity using hashing mechanism [50]. These layers use generative AI, DL, and ML approaches as the underlying hardware and software components are AI-enabled to server metaverse AI applications. The four layers are decoupled for flexibility where each layer can evolve independently.

The realization of the Metaverse depends on a massive amount of data generated using wearable and IoT devices. This data can be categorized into 1) Big Stream transient data, such as realtime localization captured from GPS, and 2) Big Streams persistent data, such as digital twin replicas including avatars, digital sensors, and digital space stored in a cloud data center. Metaverse applications require realtime decision-making and motion rendering for immersive and user-centric QoE that eliminates the line of difference between the physical and virtual worlds. In addition, in the context of the collaborative Metaverse, there is a need for strict security and privacy requirements, adding overhead on how to identify avatars and how to process and transmit users' sensitive information over the network.

These requirements in terms of ultra-low latency, high reliability, high data rates, security, and privacy make Metaverse applications bandwidth-compute-storage-hungry. Despite cloud computing can handle such processing and storage requirements for massive data, it is unable to guarantee the QoS (e.g., latency) and QoE (e.g., visual quality) requirements of applications. Consequently, edge computing is introduced to process applications close to the users to provide a realtime immersive experience. However, directing all users' requests from a Metaverse application to be processed by the edge may lead to overhead. Consequently, AI-based distributed computing and task offloading in an edge and cloud-integrated system has been introduced [51] which is used to satisfy the QoS, QoE, and interoperability requirements of Metaverse applications.

## 6. Metaverse Development Platforms

Several platforms have been developed to aid in the creation of Metaverse applications. Though these platforms are still in their infancy, they are a step forward to the realization of Metaverse applications. We divide these platforms into 3 categories: 1) gaming, 2) social media, and 3) open platform to create new applications. Most of the advancement in the Metaverse so far is done in the domain of online gaming, where entertainment and socializing among peers take place [52]. Table 3 presents a comparison between the Metaverse development platforms.

As stated in the table, the majority of the development platforms support immersive experiences supporting the creation of avatars [9], [10], [53]–[57]. However, among those platforms, the Horizon workrooms development platform [9] does not provide supporting documents that would aid the developers. In addition to the creation of avatars, the development platforms [10], [53]–[55], [57] involve interactive 3D VR and AR, visual editor/drag and drop functionality, realistic physics simulation, and visual and interactive elements. These platforms run on top of multiple operating systems and devices. In particular, [53] runs on Windows, macOS, Linux, iOS, and Android operating systems, and on Xbox, PlayStation, and Nintendo Switch devices, [10] on the Microsoft Windows operating system and Oculus Quest device, [54] on Windows operating system, and on VR headset devices including Oculus Rift and HTC Vive, [55] on Windows, Linux, macOS, Classic Mac OS operating systems, and [57] on Android, iOS operating system, and VR devices. A report published in July 2023 stated that Roblox [58] is one of the most used Metaverse development platforms with over 56 million daily active users [59]. Regarding security, Gather [60] stores data in a cloud using encryption techniques, whereas the Unreal engine uses blockchain functionalities to ensure security [61].

**Table 3:** Metaverse development platforms.

| Platform name | Company name | Development year | Characteristics | | | | | Types of apps | Interactive apps | Immersive experience (Avatars) | Supporting platforms | Supporting documents | Supporting languages |
|---|---|---|---|---|---|---|---|---|---|---|---|---|---|
| | | | Interactive 3D VR and AR | Visual editor/ Drag and drop | Realistic physics simulation | Visual and interactive elements | Security (encryption/ blockchain) | | | | | | |
| Unity [53] | Unity Technologies, Denmark | 06/2005 | ✓ | ✓ | ✓ | ✓ | ✗ | Gaming and generic | ✓ | ✓ | Windows, macOS, Linux, iOS, Android, Xbox, PlayStation, Nintendo Switch | ✓ | C#, JavaScript, and Boo |
| Horizon workrooms [9] | Facebook (Meta), United States | 08/2021 | ✓ | ✓ | ✓ | ✓ | ✗ | Social Media, collaborative meetings | ✓ | ✓ | Android OS | ✗ | N/A |
| Horizon Wolds [10] | Meta, United States | 12/2021 | ✓ | ✓ | ✓ | ✓ | ✗ | Gaming | ✓ | ✓ | Microsoft Windows and Oculus Quest | ✓ | N/A |
| Sansar [54] | Linden Lab, San-Francisco | 07/2017 | ✓ | ✓ | ✓ | ✓ | ✗ | Gaming and social media | ✓ | ✓ | Windows and VR headsets including Oculus Rift and HTC Vive | ✓ | C# |
| Gather [60] | Gather Presence Inc | Not Reported | ✗ | ✓ | ✗ | ✗ | ✓ | Collaborative meetings and remote classrooms | ✓ | ✗ | Windows and macOS | ✓ | Not applicable |
| Second Life [55] | Linden Lab, San-Francisco | 06/2003 | ✓ | ✓ | ✓ | ✓ | ✗ | Gaming and social media | ✓ | ✓ | Windows, Linux, macOS, Classic Mac OS | ✓ | Linden scripting language, Python, C++, XML |
| Enjin [62] | Enjin Pte. Ltd | 2009 | ✓ | ✓ | ✗ | ✓ | ✗ | Gaming and generic | ✓ | ✗ | Windows, Linux, and macOS | | Python, Java and C++ |
| Stageverse | Stageverse | 2017 | ✓ | ✓ | ✗ | ✓ | ✗ | Gaming and generic | ✓ | ✓ | Android, iOS, and Meta Quest 2 | ✓ | Not applicable |
| Roblox [58] | Roblox Corporation | 09/2006 | ✓ | ✓ | ✗ | ✓ | ✗ | Gaming | ✓ | ✗ | Windows, Android, iOS, Xbox series X and S, macOS, and Fire OS | ✓ | Lua |
| Unreal Engine [61] | Epic Games | 1998 | ✓ | ✓ | ✗ | ✓ | ✓ | Gaming | ✓ | ✗ | Windows, Linux, and macOS | ✓ | C++ |
| Rooom [57] | Rooom AG | 2018 | ✓ | ✓ | ✓ | ✓ | ✗ | Generic | ✓ | ✓ | Android, iOS, and VR | ✓ | Not reported |

## 7. A Taxonomy of Metaverse Applications

Metaverse is a user-centric computing paradigm that enables immersive virtual and physical interactions for users and their avatars in dynamic, realtime, immersive, and interactive smart city applications. It opens up opportunities for a thriving economy in a world where people can game, shop, meet, work, and learn from each other, all from their physical location. Metaverse goes beyond VR and AR by making applications' contextual contents persistent, whereas users can go out and in Metaverse to resume collaboration and transactions that have been taking place. Figure 7 shows a taxonomy of Metaverse applications. We categorize these applications into education, smart healthcare, smart mobility, gaming and entertainment, business, real estate, social media, and manufacturing.

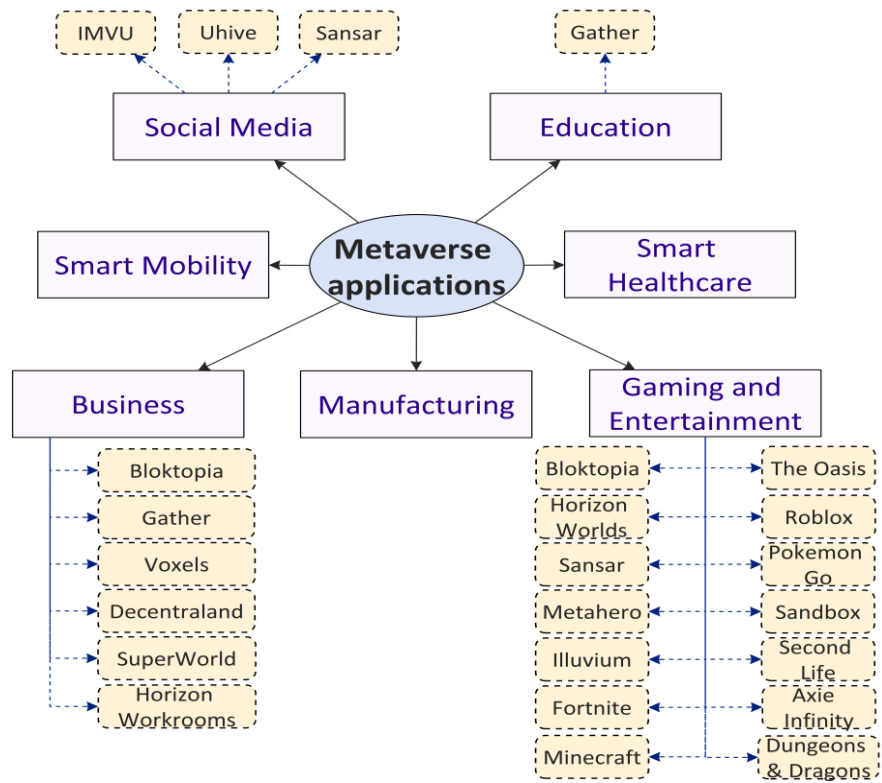

**Figure 7:** Taxonomy of Metaverse applications.

- *Education:* Metaverse can enhance the learning experience by allowing teachers to create 3D surroundings wherein students can interact with virtual objects in space to understand different concepts such as biology, anatomy, cosmology, and geometry. education for students, for example, the ability to inspect 3D objects, moving and rotating virtual objects in space to understand certain concepts, important in topics like biology, anatomy, cosmology, and geometry. Furthermore, students can simulate 3D space to conduct scientific experiments in different scenarios and analyze their results from different angles. In addition, Metaverse can aid in training professionals using 3D models in digital space will be effective, efficient, and safer compared to training them on physical machines.
    - Gather developed by Gather Presence Inc. [60] – It enables users to create virtual classrooms for education without an immersive experience. In Gather virtual classroom, students and faculty are represented using 3D characters that can move around the digital space. Students can talk to each other and collaborate on projects, and faculty can hold open office hours and talk with students pre or post-lectures.
- *Smart healthcare:* Metaverse can revolutionize the healthcare industry by providing personalized patient-centric care regardless of patients' and healthcare providers' locations. Furthermore, 3D models can aid in more accurate diagnostics and surgeries as doctors can examine a patient from different angles in the digital world. Metaverse can ensure the security and privacy of patients' sensitive information by storing and sharing data in virtual environments using blockchain. To our knowledge, there is no deployed application in this category so far.

- *Smart mobility:* Metaverse can reshape transportation by the concept known as MetaMobility in which smart devices or robots will interact with users to provide mobility services. It can aid in traffic management and parking by allowing users to view 3D models of different locations in realtime. In addition, the Metaverse can allow the creation of intelligent vehicles by creating and simulating 3D models before production. To our knowledge, there is no deployed application in this category so far.
- *Gaming and entertainment:* Metaverse-based virtual ecosystems connect gamers in a virtual shared space to interact, play games, and socialize. It gives users a multisensory immersive experience while gameplay experience. Furthermore, it allows users to create their subgames in 3D space.
    - Second Life developed by Linden Lab [55] – Is an online multiplayer gaming platform for an immersive experience that enables users to create representative avatars and interact with other users and content created by other users within a multiplayer digital world. It simulates a free market economy as players can buy and sell virtual goods with virtual money.
    - Sandbox developed by Animoca Brands [63] – Is an online gaming platform that enables users to create games, monetize games and play games created by other users.
    - Roblox developed by Roblox Corporation [58] – Is an Ethereum-based online gaming platform that enables users to program games and play games created by other users. Players can use Robux, virtual currency, to make in-app purchases.
    - The Oasis [64] – Is an online multiplayer gaming platform that harnesses the power of NFTs and enables the creation of multiplayer social games with decentralized finance.
    - Minecraft developed by Mojang Studios, Xbox Game Studios, Telltale Games, 4J Studios, Double Eleven, and Other Ocean Interactive [65] – It is an online multiplayer gaming platform for an immersive experience that enables users to create and share their virtual world online.
    - Pokémon Go developed by Niantic, Inc. [66] – Is an AR-based multiplayer gaming platform for an immersive experience that uses a mobile device's GPS to locate, capture, train, and battle virtual creatures known as Pokémon.
    - Axie Infinity developed by Sky Mavis [67] – Is an NFT-based online video game that enables users to play games and do in-game purchases using Ethereum-based cryptocurrencies. Users can collect NFTs that represent digital pets in Axie Infinity knowns as Axies and can perform battle between Axies collected by other users in the game.
    - Fortnite developed by Epic Games and People Can Fly [68] – Is an online gaming platform that enables users to create games and play games with other users.
    - Illuvium developed by Illuvium Decentralized Autonomous Organization [69] – It is an interoperable blockchain game developed on the Ethereum blockchain. Users in Illuvium are immersed in a 3D digital world where they explore and collect digital beasts called Illuvials. Illuvials collected by players are

- represented with NFTs on the blockchain, and they are used to battle other players to win ether cryptocurrency.
  - Metahero developed by Metahero [70] – It is a 3D scanning technology that digitally renders a real-world object based on collected appearance data. Metahero uses NFTs smart contracts to enable the creation of avatars. Users can use their avatars to play immersive games.
  - Sansar developed by Linden Lab and Metaverse Investment Ltd. [54] – It is a social virtual reality platform where users are represented using avatars with speech-driven facial emotions and motion-driven body animations. Sansar allows users to design interactive and immersive games for VR and desktop, and play games created by other users.
  - Horizon Worlds developed by Meta [10] – Is a VR online gaming platform that enables users to create online virtual games and move and interact with other players in the virtual world. The game uses full 3D motion and can be played using Oculus Rift S or Meta Quest 2 VR headset.
  - Bloktopia developed by Bloktopia [71] – Is a platform designed as a decentralized 21 floors VR skyscraper that allows users (represented using avatars) to play games and earn revenue.
  - Dungeons & Dragons by Gary Gygax and Dave Arneson [72] – Is a series of online games which allow avatars to adventure into interconnected virtual worlds with their regulations and recompense.
- **Business:** Metaverse can enable users to create virtual 3D stores where customers can remotely browse and order products. Furthermore, it can allow customers to with brands more personally leading to trust. Virtual worlds can aid in creating interactive marketing advertisements that enable users to explore and interact with products digitally. Metaverse can also be useful for remote sales and meetings in a 3D virtual space where people from around the world can interact. It enables users to buy and sell properties virtually using NFT. Realtors can create a 3D replica of the property and potential buyers have an immersive virtual tour.
  - Bloktopia developed by Bloktopia [71] – It is a platform designed as a decentralized 21 floors VR skyscraper that allows users (represented using avatars) to learn, play, and earn revenue by purchasing, selling, or leasing virtual real estate.
  - Horizon Workrooms developed by Meta [9] – It is a virtual and immersive office that allows team members to meet, collaborate, brainstorm ideas, and share presentations. Horizon workrooms can be accessed using a Meta Quest VR headset or a web browser.
  - Gather developed by Gather Presence Inc. [60] – It enables users to create virtual headquarters for remote teams without an immersive experience. In Gather virtual headquarters, team members are represented using 3D characters that can move around the digital space, communicate with other team members, and schedule meetings. It supports the integration of Slack, Google/Outlook calendar, and Outlook.
  - Decentraland developed by Decentraland Foundation [73] – Is a 3D virtual world where users can buy virtual real estate plots as NFTs using the MANA cryptocurrency based on the Ethereum blockchain. Furthermore, Decentraland enables designers to create and sell clothes and accessories for the avatars to use in the digital space.
  - SuperWorld developed by SuperWorld [74] – Is a virtual estate marketplace where a virtual world in AR is digitally mapped over the earth. SuperWorld consists of approximately 64.8 billion blocks of virtual estate that users can buy or sell using a crypto wallet. Each virtual real estate transaction is recorded in the blockchain.
- o Voxels developed by Nolan Consulting Limited [75] – It enables users to build, explore, and buy digital arts and NFTs using the Ethereum blockchain. In addition, Voxels allows users to create costumes for the avatars and make friends in the digital space.
- *Social media:* Metaverse can allow users to interact with each other in an immersive and lifelike manner using avatars.
    - o IMVU developed by IMVU [76] – Is an online and immersive social network that enables users to create 3D avatars, connect to other users, and chat with different users around the world.
    - o Sansar developed by Linden Lab and Metaverse Investment Ltd. [54] – Is a social virtual reality platform where users are represented using avatars with speech-driven facial emotions and motion-driven body animations. Sansar allows users to have conversations in VR, watch videos, and play games with each other. Enables users to create virtual interactive sessions and events.
    - o Uhive developed by Uhive [77] – It enables users to create virtual spaces that will allow them to share their content in digital space, view content created by other users, and interact with other users. It includes Uhive Token (HVE2), a native cryptocurrency of the Uhive social network, that allows users to buy virtual spaces, reward content creators, send tips or donations to other users, trade on cryptocurrency exchanges, and make in-app purchases.
- *Manufacturing:* Metaverse can be used to test products under different scenarios and perform predictive maintenance. It can play the role of assistant for workers in the field to manipulate objects virtually before the physical repair aiding in saving time. It can include digital twins for machines in the field to repair equipment across the globe using 3D virtual space without human intervention in the field. To our knowledge, there is no deployed application in this category so far.

## 8. Open Challenges and Research Directions

The proposed realtime scalable-centric vision comprises an architecture that is user-centric and enables different users and avatars to interact in the Metaverse framework. It enables interaction in a way to increase the QoS and satisfy the SLA requirements. The framework promotes scaling up to meet the heterogeneous and dynamic requirements of Metaverse applications. The open challenges to the Metaverse framework include the following:

- **Realtime:** Metaverse applications require ultra-low latency response for realtime virtual world rendering to provide multiple users with seamless and immersive QoE. These applications have stringent network bandwidth and ultra-low latency requirements for realtime responses. For instance, a latency of less than 100 ms is required for realtime gaming [78]. The latency requirements become more stringent with an increasing level of interaction with a Metaverse application. One of the solutions to achieve low latency is to process a portion of the Metaverse application on close by edge servers whereas the computation-hungry scene rendering can be performed in the cloud. Huynh et al. [79] proposed to optimize offloading portion, edge caching policies, bandwidth allocation, and computation resources to guarantee stringent low latency requirements of the digital twin-enabled Metaverse. Yu, Chua, and Zhao [80] proposed a multi-agent reinforcement learning approach to ensure low latency in Metaverse by optimizing computation offloading, transmission power, and channel allocation decisions.

- **Scalability:** In the Metaverse market, the number of users is expected to amount to 1,461.00m users by 2030 [81]. The applicability of Metaverse for scenarios that require the creation of multiple avatars simultaneously, for instance, in a virtual conference, might be limited. This is because, with an increasing number of avatars, there will be a communication and computation overhead leading to violations of Metaverse's stringent low latency and high bandwidth requirements. Consequently, the scalability of the Metaverse is a major issue. [82] found that Workrooms, a Metaverse platform, may suffer from scalability issues when there are more than 10 participants. This is due to an increase in downlink bandwidth requirement. This leads to communication overhead which puts a burden on the network scalability. To address the scalability issues, several solutions have been proposed in the literature. One of the most prominent approaches to use is peer-to-peer communication and computing model [83] where data will be stored and processed on multiple network-wide peer servers and then combined on a centralized server to render the virtual world [82]. However, rendering multiple avatars on a centralized server will still be resource hungry. This issue could be further addressed by performing rendering on remote cloud servers with higher computing capabilities [84]. Consequently, the entire scene with multiple users will be rendered on remote servers and only the scene will be transmitted to the client. However, data transmission from the cloud servers to the client requires ultra-low latency. Furthermore, the scalability of the blockchain, trading, and privacy-preserving layer for the Metaverse is a major issue. A blockchain network grows rapidly in terms of participants and data, leading to an increasing number of transactions and block validations. One of the notable approaches is to implement a lightweight peer-to-peer blockchain network is divided into clusters based on the geographical locations of network participants [85]. A copy of the ledger is maintained per cluster by the cluster head. The scalability of a blockchain network can be further improved by using a non-encapsulated integrated blockchain-cloud architecture where the data from the physical world, used to render a virtual world, is stored in a cloud database while the metadata, such as data hash, scene rendering events, and access control policies, is recorded in blockchain ledge for security and privacy [86].
- **High energy consumption:** Metaverse involves IoT devices, edge data centers, and cloud data centers to deploy learning, prediction, and rendering models. These devices and data centers consisting of thousands of servers consume a high amount of energy. It is predicted that by 2025, data centers will consume 4.5% of the total global energy consumption [87]. High energy consumption leads to increased electricity costs and global warming due to carbon emissions. It is estimated that by 2040, Information and Communications Technology (ICT) will account for 14% of global carbon emissions [88]. Possible solutions include introducing energy-efficient hardware components [89] and/or energy-aware resource provisioning strategies in edge and cloud data centers [48].
- **Resource provisioning to optimize QoS and energy consumption:** Metaverse is expected to be accessed by multiple users that might be in different parts of the world. Rendering of this huge amount of data is computationally expensive and requires critical provisioning of edge and cloud resources to optimize energy consumption and Metaverse QoS. For instance, when a user is mobile, the latency between the user's motion and its avatar, which is perceived by other users, is a crucial QoS metric to optimize. One of the important approaches is devising algorithms, to efficiently allocate resources in a distributed and cloud computing for Metaverse applications, that can be evaluated using formal modelling [90]. In addition, several works in literature have proposed QoS-aware [91], [92] resource provisioning and energy-aware [48], [93] approaches in an integrated IoT, edge, and cloud computing environment that aid in the realization of seamless Metaverse applications while optimizing the energy consumption of the edge and cloud infrastructure alongside the performance of IoT applications with stringent requirements.

- **Cost and complexity:** The complexity of implementing and deploying a Metaverse application, and the associated cost for hardware components and devices are major obstacles to the adoption of the technology. For instance, the cost of Meta Quest Pro, a hardware component, costs up to 1000 USD [94]. Cost-effective hardware devices and simpler implementation solutions should be developed to mitigate this issue.
- **Security and data privacy:** Metaverse requires users to provide their identification as well as biometric information to access headset devices, leading to security and privacy concerns. The traditional secure communication protocols, such as TLS and DTLS, will no longer be enough to ensure data privacy. Furthermore, the Metaverse will highly rely on Digital Twin technology to communicate between physical and virtual worlds. If learning models, supporting digital twins, are attacked in the virtual world, then the consequences in the physical world will be a threat to security and data privacy. One of the possible solutions to ensure the security and privacy of users' information and virtual models is the use of blockchain [16]. In addition, avatars and digital twins can hide identities leading to inequalities and biases, making it challenging to achieve fairness in virtual worlds.
- **The need for governance against abuse:** Harassment and bullying are major concerns in the Metaverse that require governance, as this may lead to mental health issues. Metaverse is not only limited to text or voice-based bullying, such as in social media but promotes body movement-based harassment through avatars [95], [96]. To address this issue, Meta introduced a safety feature called Safe Zone, which allows users to activate a protective bubble when feel threatened, which will not allow other users to talk, touch, or interact with the safe zone-enabled user [97]. However, existing solutions are either blocking/muting the user who is bullying/harassing or reporting community standard violations. There is a lack of governance towards content moderation that should be addressed.
- **Standardization and interoperability:** With increasing attention toward Metaverse, various companies, developers, and organizations are developing their Metaverse platforms. These platforms have disparate architectures, and programming languages, and use different hardware components and IoT devices. Consequently, it becomes difficult to create avatars using data from different platforms and devices. Standardization will help organizations to develop platforms that allow interoperability. To address this issue, initiatives such as Metaverse Standards Forum are established [98].
- **Health-related risks:** An extensive use of Metaverse can lead to physical and/or mental health issues [99]–[101]. Physical health conditions include motion sickness, accidents due to collisions with nearby objects, and eye fatigue [99], [100]. Mental health problems, such as depression and anxiety can be caused by a lack of physical interactions, or dissociation from reality [101].

## 9. Summary and Conclusions

Metaverse is an online immersive 3D virtual environment where people via their avatars can interact with digital objects as if they were real. Unlike VR/AR platforms which focus on the development of monolithic centralized applications, the Metaverse platform enables the development of distributed VR/AR applications, giving rise to virtual 3D immersive social interactions between avatars representing humans. The evolution of the design of the upcoming generation of networks and mobile systems will depend on the innovation of the users in designing new applications. Metaverse is an ideal emerging technology to influence this domain by providing new types of immersive interactivity, dynamic and evolving data, and the required computation resources for creating revolutionary applications. With Metaverse, the Internet goes beyond being a facilitator of the exchange of information and ideas among users, to being an enabler for sharing virtual objects and 3D immersive communication in a peer-to-peer network in realtime.

In this paper, we presented a taxonomy of Metaverse applications and their limitations to shed light on smart city applications domains where creative Metaverse applications should be developed. The proliferation of development platforms motivates us to provide a comparison among the existing platform along with their features and limitations for the developers. While AI is an enabler for developing immersive Metaverse applications, it is also a driving force to provide intelligent solutions for realtime and scalable Metaverse as highlighted by our proposed layered architecture. Empowering edge and cloud computing data centers with AI-based solutions is crucial to efficiently schedule Metaverse requests and dynamically provision the necessary resources for increasing the Metaverse QoS, and QoE, and satisfying its SLA requirements. In proposing a new framework for Metaverse applications, we highlighted the associated challenges, ranging from realtime, scalability, provisioning, and high energy consumption, to cost and complexity, data privacy, need for governance, standardization and interoperability, and health-related risks.

**Acknowledgement**
The first author would like to thank the HCI Lab team members of the School of Computing and Information Systems of the Faculty of Engineering and Information Technology at The University of Melbourne for the Lab tour of VR equipment and applications.

This research is funded by the National Water and Energy Center of the United Arab Emirates University (Grant 12R126).